\documentclass[conference,10pt]{IEEEtran}

\IEEEoverridecommandlockouts
\usepackage{comment}

\usepackage{booktabs}
\usepackage[utf8]{inputenc}
\usepackage[ruled,vlined]{algorithm2e}
\usepackage{microtype}
\usepackage{graphicx}
\usepackage{paralist}
\usepackage{tabularx}
\usepackage{balance}
\usepackage{multicol}
\usepackage{multirow}
\usepackage{pbox}
\usepackage{enumitem}
\usepackage{colortbl}
\usepackage{pifont}
\usepackage{xspace}
\usepackage{url}
\usepackage{tikz}
\usepackage{rotating}
\usepackage{balance}
\usepackage{float}
\usepackage[TABBOTCAP]{subfigure}
\usepackage{tcolorbox}
\usepackage{stackengine}
\usepackage{paralist}

\usepackage[hidelinks]{hyperref}

\makeatletter
\newcommand{\linebreakand}{%
  \end{@IEEEauthorhalign}
  \hfill\mbox{}\par
  \mbox{}\hfill\begin{@IEEEauthorhalign}
}
\makeatother

\def\BibTeX{{\rm B\kern-.05em{\sc i\kern-.025em b}\kern-.08em
    T\kern-.1667em\lower.7ex\hbox{E}\kern-.125emX}}

\usepackage{listings}

\input{macro}

\newcommand{\toolName}{\textsc{CodeGenLink}\xspace}
\newcommand{\links}{LinkSearcher\xspace}
\newcommand{\codegl}{CodeGenLink\xspace}

\begin{document}

%\title{Where is Code Generated by LLMs Coming From? A Study with Gemini and Bing CoPilot}
%\title{Where is Code Generated by LLMs Coming From? A Study with Gemini and Bing CoPilot}
\title{CodeGenLink: A Tool to Find the Likely Origin and\\License of Automatically Generated Code}
%\title{CodeGenLink: A Tool Aiding Developers \\ to Find the Likely Origin and License \\ of Automatically Generated Code}

\author{
\IEEEauthorblockN{Daniele Bifolco}
\IEEEauthorblockA{\textit{University of Sannio, Italy}\\
d.bifolco@studenti.unisannio.it}
\and
\IEEEauthorblockN{Guido Annicchiarico}
\IEEEauthorblockA{\textit{University of Sannio, Italy}\\
g.annicchiarico@studenti.unisannio.it}
\and
\IEEEauthorblockN{Pierluigi Barbiero}
\IEEEauthorblockA{\textit{University of Sannio, Italy}\\
p.barbiero3@studenti.unisannio.it}
\linebreakand 
\IEEEauthorblockN{Massimiliano Di Penta}
\IEEEauthorblockA{\textit{University of Sannio, Italy}\\
dipenta@unisannio.it}
\and
\IEEEauthorblockN{Fiorella Zampetti}
\IEEEauthorblockA{\textit{University of Sannio, Italy}\\}
fzampetti@unisannio.it}


\maketitle
\thispagestyle{empty}

\begin{abstract}
Large Language Models (LLMs) are widely used in software development tasks nowadays. Unlike reusing code taken from the Web, for LLMs' generated code, developers are concerned about its lack of trustworthiness and possible copyright or licensing violations, due to the lack of code provenance information.
This paper proposes \toolName, a GitHub CoPilot extension for Visual Studio Code aimed at (i)~suggesting links containing code very similar to automatically generated code, and (ii)~whenever possible, indicating the license of the likely origin of the code.
\toolName retrieves candidate links by combining LLMs with their web search features and then performs similarity analysis between the generated and retrieved code.
Preliminary results show that \toolName effectively filters unrelated links via similarity analysis and provides licensing information when available.

Tool URL: \url{https://github.com/danielebifolco/CodeGenLink}

Tool Video: \url{https://youtu.be/M6nqjBf9\_pw}
\end{abstract}


\section{Introduction}
\label{sec:intro}
Large Language Models (LLMs) are nowadays widely used to perform several development tasks, ranging from code generation, testing, re-documentation, or support to the software development process~\cite{fan2023large,hou2023large,msr2024}.

Considering code generation, \ie the most prominent among the previously mentioned tasks, there is one key element that makes LLM promising and worrisome at the same time, when compared to code reuse from the Web. For the latter, ownership can often be inferred, enabling an assessment of the source’s trustworthiness. However, this is not always possible for LLM-generated code. Furthermore, reuse and redistribution of LLM-generated artifacts can lead to license or copyright violations. \revised{Yang \etal~\cite{yang2024unveiling} were the first assessing the ``memorization'' effect within pre-trained models, \ie models memorizing and producing
source code verbatim. The reuse of this code may lead to copyright violations, as well as to issues with code attribution~\cite{colombo2025possibility, KatzyPDI24}. Specifically, Colombo \etal~\cite{colombo2025possibility} found that more context increases ChatGPT's reproduction of copyleft code, while higher temperature reduces this risk. Last but not least, researchers have also looked at the reuse of code from training
sets~\cite{abs-2402-09299,xu2025licoevalevaluatingllmslicense}}. 

In a previous work \cite{msr2025}, we analyzed the extent to which LLMs, if properly prompted, could provide relevant links to the code they generate. However, although LLMs generate multiple links, only a few are likely to be the ``true origin'' for the generated code.

In this paper, we propose \toolName, a tool integrated in GitHub Copilot~\cite{github_copilot} \revised{(hereafter referred to as Copilot)} as a Visual Studio \revised{(VS)} Code extension, that assists developers in determining the likely origin (when available) of the generated code, and the license under which it should be redistributed. The tool has two operational modes:
\begin{compactenum}
\item \textbf{Direct code + link retrieval (Mode 1):} The user, through the Copilot chat, asks the LLM to generate a code snippet. The LLM output is automatically used by \toolName to retrieve links similar to the automatically generated code by enabling the web search. 
\item \textbf{Link retrieval (Mode 2):} The user selects the automatically generated code directly within the IDE and asks \toolName to retrieve relevant links for it. Note that in principle this mode can be used to seek the provenance of any code snippet, even if it is not automatically generated.
%When using GitHub Copilot, the user can receive support for code generation directly within the IDE. In this case, it is still possible to select the automatically generated code and . 
\end{compactenum}

%\revised{Depending on the chosen operational mode, \toolName interacts with the LLM in different ways to retrieve relevant links.} Those links are then processed to extract the code snippets, which are subsequently compared with the automatically generated code (for Mode 1) or with the selected code from the IDE (for Mode 2). 
\toolName combines text similarity and clone detection to identify likely code origins, then inspects each relevant link to extract associated license information.
%The comparison is performed using both text similarity and clone detection to retain only the links most likely to be the origin of the generated code. Furthermore, \toolName examines the content of each potentially relevant link to determine the license associated with the code.

Unlike tools such as LiCoEval~\cite{xu2025licoevalevaluatingllmslicense}, which use code similarity analysis to check whether generated code is part of an LLM’s training set, \toolName 
\revised{ searches the Web for code snippets that match the generated code and identifies their associated licenses, allowing it to work even when the LLM's training data is inaccessible.}

We have performed a preliminary evaluation of \toolName using  101 coding tasks from \textsc{CodeSearchNet}~\cite{husain2020codesearchnetchallengeevaluatingstate} and 100 coding tasks from \textsc{CoderEval}~\cite{DBLP:conf/icse/YuSRZZMLLWX24}. The evaluation results are promising since, even if LLMs do not always provide relevant links, the similarity mechanism filters them out and only provides the users with the ones that are ``likely'' the origin of the code.

%\toolName is available for download under the Apache 2.0 License, and we plan to publish it on the VS Code extension marketplace.

\section{Architecture}
\label{sec:architecture}
\begin{figure}[t]
    \centering    \includegraphics[width=\linewidth]{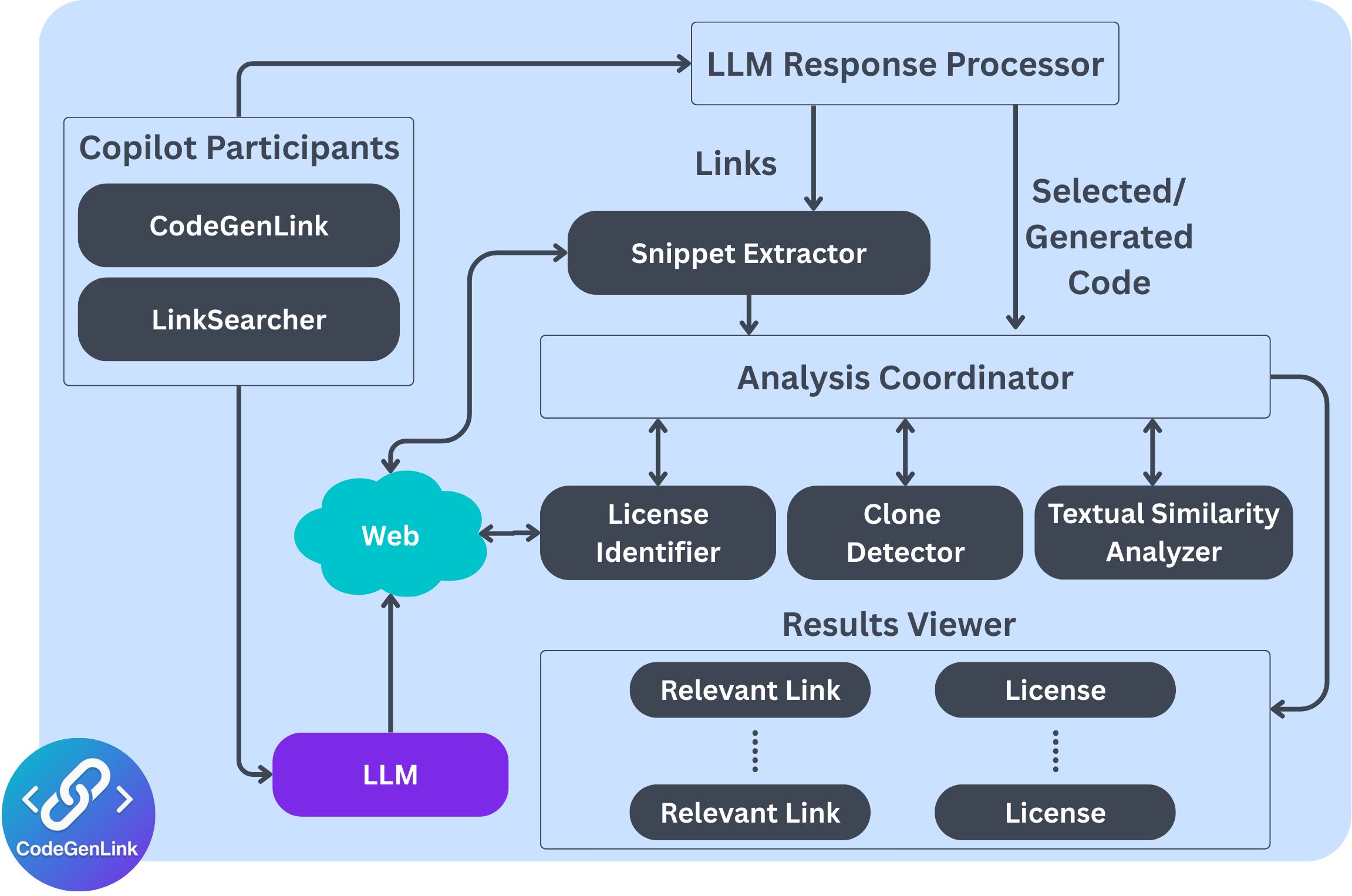}
    
    \caption{\toolName Architecture}
    \vspace{-3mm}\label{fig:architecture}
    \vspace{-2mm}
\end{figure}

\figref{fig:architecture} depicts the \toolName's architecture. %\toolName has been developed as a GitHub CoPilot extension for VS Code. 
\toolName integrates with Copilot through the Chat Participants~\cite{chatextensionvscode} interface provided by VS Code. A participant is an agent interacting with the LLM, referred to in the chat by prepending its name with a ``@". \toolName has two participants as entry points supporting the two \revised{operational modes} outlined in the introduction,  \ie \textbf{\textit{CodeGenLink}}~\revised{(\textbf{Mode 1})} and \textbf{\textit{LinkSearcher}}~\revised{(\textbf{Mode 2})}.
To provide links, \toolName relies on an LLM with web search capabilities. We used OpenAI’s \textit{web-search-preview}~\cite{openai_websearch}, though other LLM APIs with similar support can be used. 

%Web search can be enabled by configuring the \texttt{tool} array in the API request to \texttt{\{ type: \"web\_search\_preview\" \}} \cite{openai_websearch}.
%We used the \texttt{gpt-4o-search-preview} model, but other LLMs that support web search can be plugged in as well.
%The two-participant design allows for flexible, context-sensitive interactions with GitHub Copilot, even if both rely on a shared backend for tasks such as similarity analysis and license detection. 
%The coordination of all components is managed by the \textit{Analysis Coordinator}, acting as the central controller.

%\toolName can be fully configured through the settings interface provided by VS Code: users can set a range of parameters, including the model, OpenAI API key, similarity thresholds, the minimum number of lines of code for the code snippets to be analyzed, paths to external tools (\ie clone detector and license classifier), and the output directory. 

%All core modules rely on this configuration layer to retrieve runtime parameters, making the system easily adaptable and maintainable.

%\subsection{GitHub Copilot Integration}
%\subsection{GitHub Copilot Participants}
\revised{\textit{\textbf{GitHub Copilot Participants.}}}
We developed two specialized participants to support distinct operational modes:

\begin{compactitem}
    \item \textbf{\textit{\codegl}}. This participant is invoked when the user interacts with the Copilot chat, prefixing the prompt with \texttt{@\codegl}. It represents operational \textbf{Mode 1}, where the model first generates code that meets the user's requirements. After that, the provided output is used to prompt the model again to search the Web for relevant links related to the generated code. 
    \revised{The prompt for code generation is inherited from }
    our previous study~\cite{msr2025} \revised{(details in \cite{codegenlink_repo})}.  
    %for code generation, we suggest structuring the prompt as:    
    %\begin{quote}
    %\emph{``You are a Senior $<<$LANGUAGE$>>$ developer. Then give me a $<<$LANGUAGE$>>$ code snippet about: $<<$QUERY$>>$"}
    %\end{quote}
    \revised{To retrieve relevant links from the Web, the suggested prompt is}: 
    
    \begin{quote}
    \emph{``Search the Web to find links where I can get more information about the generated code present in this response."}
    \end{quote}

    \item \textbf{\textit{\links}}. This participant is activated when the user selects a code fragment within the IDE and triggers operational \textbf{Mode 2}. It constructs a structured prompt \revised{similar to the previous one, but copying the selected code from the IDE and specifying the programming language (details can be found at~\cite{codegenlink_repo}).}%aimed at retrieving links relevant to the selected code snippet, \eg automatically generated code by GitHub Copilot within the IDE: 
    %\begin{quote}
    %\emph{``Search the Web to find links where I can get more information about this code. $<<$LANGUAGE$>>$, $<<$CODE\_SNIPPET$>>$"}
    %\end{quote}
    %The programming language ($<<$LANGUAGE$>>$) of the selected code snippet is automatically extracted using the \texttt{vscode} module, which contains the Visual Studio Code extensibility API 
    %a specific API \FIORELLA{which one} 
    %of Visual Studio Code. 
\end{compactitem}

%The LLM output, \ie automatically generated code and retrieved links, are used as input for the \textit{LLM Response Processor} component. 

%\subsection{LLM Response Processor}

%The first step involves activating the appropriate participant, depending on the desired use case: either generating code via an LLM and retrieving its related links (\textit{\codegl}), or identifying the related links for a code snippet already present in the user’s software project (\textit{\links}).
%The \textit{LLM Response Processor} 
\revised{\textbf{\textit{LLM Response Processor.}} This component }converts the unstructured LLM output, \ie automatically generated code and retrieved links, into structured data that can be used by the \textit{Analysis Coordinator} component for both the similarity analysis and the license identification. 

%\subsection{Snippet Extractor}

\revised{\textbf{\textit{Snippet Extractor.}}}
 %serves as an interface between \toolName and external web sources. Specifically, it 
This component parses the HTML content from retrieved links to extract code snippets using domain-specific strategies, due to the heterogeneity of web content structures. 
%of the retrieved links to extract code snippets from them. Given the heterogeneity of web content structures, the component employs domain-specific extraction strategies tailored to the structural characteristics of each source. 
For instance, it pulls Stack Overflow code from nested \texttt{<pre>} and \texttt{<code>} tags, and accesses GitHub code via raw URLs. \toolName currently supports popular domains such as \textit{GitHub, Stack Overflow, GeeksForGeeks}, with a design allowing easy extension of extraction rules.
%The current version of \toolName can extract code snippets from widely used domains such as \textit{GitHub, Stack Overflow, GeekForGeeks, w3schools} and \textit{php.net}, but it is designed to easily extend the set of extraction rules. 

%\subsection{Analysis Coordinator}

%The \textit{Analysis Coordinator} 
\revised{\textbf{\textit{Analysis Coordinator.}}} This component pairs code from retrieved links with selected or generated code snippets, then runs similarity checks to filter out irrelevant links and keep only those with closely matching code.
%This component creates pairs of extracted code snippets from retrieved links and selected/automatically generated code snippets. Each pair will be used for the similarity analysis to filter out ``spurious'' links and provide to the users only the links having at least one code snippet whose content is highly similar to the selected/generated one. 
The current version of \toolName allows the user to filter out irrelevant links relying on textual similarity metrics and/or clone detection techniques. Specifically, the clone detector component interfaces with \textsc{CCFinderSW}~\cite{8305997} to identify potential code clones between paired snippets. The textual similarity analyzer uses a vector-based approach: it tokenizes each code snippet, converts it into a vector, and computes cosine similarity between vectors.

%This analysis yields a metric known as \textit{clone ratio}, which quantifies the degree of structural similarity between the two code fragments. 
%The textual similarity analyzer leverages a vector-based textual comparison technique. It tokenizes each code fragment and transforms it into a vector representation. These vectors are subsequently used to compute the cosine similarity.

%\subsection{License Identifier}

%The \textit{License Identifier} 
\revised{\textbf{\textit{License Identifier. }}This component } aims at detecting the license to be used when reusing/redistributing a code snippet taken from the Web. To this aim, \toolName supports several detection strategies. 
%Specifically, for links pointing to GitHub, the module first queries the GitHub API to retrieve repository metadata. If no license is found in the metadata, it then searches for a \texttt{LICENSE} file within the repository. If such a file is found, we rely on the results of a classification done by an external tool, \ie the \textsc{Google’s License Classifier}~\cite{googlelicenseclassifier}. The output is a standard SPDX \cite{spdx2025} license identifier.
%\FIORELLA{how?};
%For other types of links, \toolName scans the HTML content of the page to search for the presence of license-related keywords. These keywords are defined in a hardcoded list. The module checks for matches against this list to infer the possible license under which the code may be reused. 
%\FIORELLA{it is unclear when this tool is invoked} The output is a standard SPDX \cite{spdx2025} license identifier.
For URLs pointing to GitHub repositories, the component first queries the GitHub API to obtain metadata, including license information. If no license is found in the metadata, the component searches the repository for a \texttt{LICENSE} (or similar names) file and, if found, analyzes it using an external license classification tool, \ie \textsc{Google’s License Classifier}~\cite{googlelicenseclassifier}, to identify the license. 
In other cases, the license is determined based on the policy of the source website. For instance, all code snippets on Stack Overflow are licensed under the CC BY-SA 4.0 license. To do this, \toolName maintains a list of domain/license pairs.
%while content from PHP.net is released under the PHP License v3.01. Since these platforms apply a uniform license to all their content, \toolName directly assigns the corresponding license identifier without further analysis. For websites such as GeeksforGeeks and W3Schools, where no explicit license is uniformly declared, \toolName provides the URL of the original page. This is necessary to fulfill attribution requirements and to allow users to manually verify the applicable reuse conditions.
For other links, \toolName scans the webpage’s HTML to search for license keywords from a predefined SPDX-based list~\cite{spdx2025} and uses any matches to infer the code's license.

%For the other links, \toolName analyzes the HTML content of the target webpage, searching for license-related keywords, drawn from a predefined list obtained from SPDX~\cite{spdx2025}. Any matches found are used to infer the potential source code license.

%Once the license is determined, the result is then mapped to a standard SPDX license identifier~\cite{spdx2025} and produced as output. 

%\subsection{Results Viewer}

\revised{\textbf{\textit{Results Viewer. }}}Once all similarity and license data are collected, a filtering step is applied. Based on thresholds defined in the configuration, only code pairs exceeding a given \textit{similarity threshold} are retained. This helps to reduce noise and ensures that only potentially relevant links are presented to the user.
The results are aggregated and shown using a WebView, showing a table containing the URL(s) of the likely origin of the generated code snippet, with its related license information, if any.
Last, the component is responsible for presenting the results inside the IDE. 
%It generates a dynamic HTML view that displays each accepted code pair and the corresponding source URL and license information, within VS Code.

\subsection{\toolName Limitations}

%\revised{\textbf{\textit{\toolName Limitations. }} }
\toolName has limitations we acknowledge:
\begin{compactitem}
\item It provides evidence of similarity between generated code and code on the Web, not a guarantee that the generated code originates from the provided URL. Also, it currently does not tell which one was generated by the AI.
\item Its performance may depend on the Web search capability of the used LLM and, to a minor extent, on the used clone detector and licensing classifier.
\end{compactitem}

\section{CodeGenLink in Action}
\label{sec:use_scenario}

To use \toolName, first, we must install the extension in VS Code and set all the parameters needed to use it successfully. The installation procedure, detailed in the Repository README \cite{codegenlink_repo}, requires the installation of Python dependencies and external tools (\textsc{CCFinderSW} and \textsc{Google License Identifier}).

\begin{figure}[htb]
    \centering
    \includegraphics[width=\columnwidth]{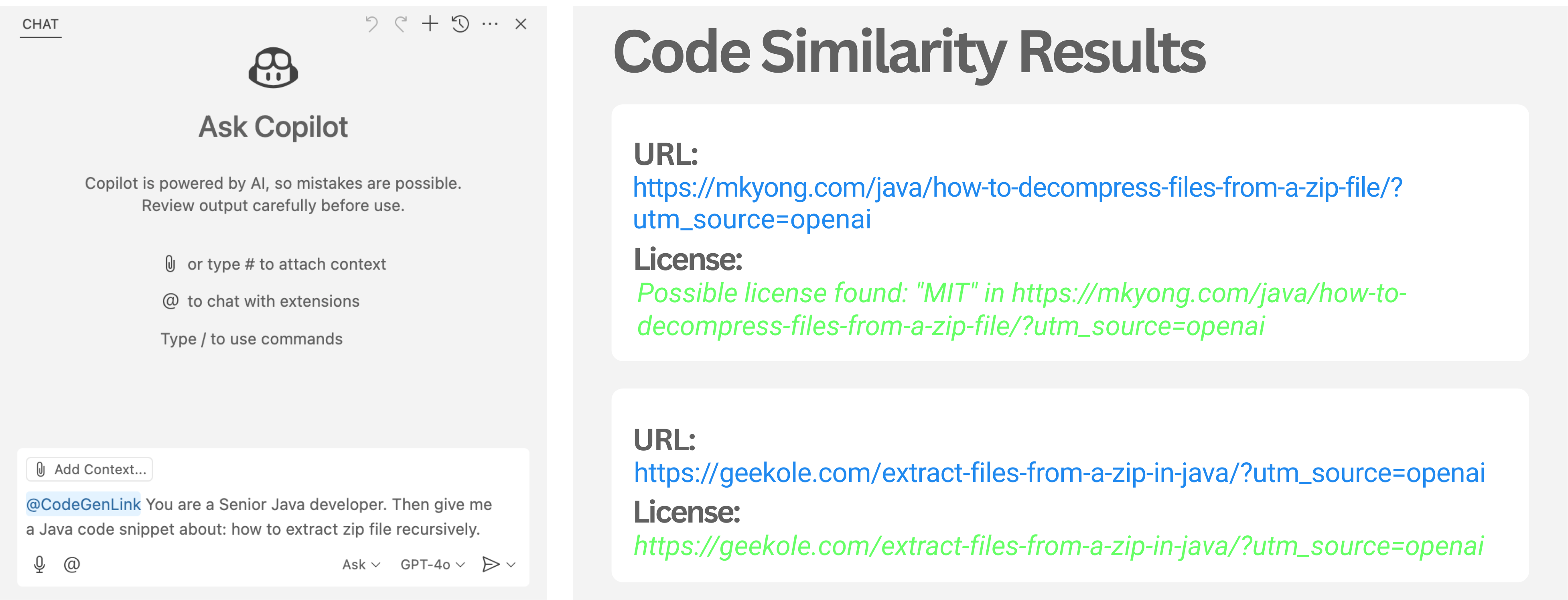}
    \caption{Usage example of the @\codegl participant}
    \label{fig:screen_prompt}
    \vspace{-2mm}
\end{figure}

\toolName supports developers in their coding activities by providing two operational modes: (i) via Copilot Chat using the \texttt{@\codegl} participant, or (ii) by selecting code directly within the IDE using the  \texttt{@\links} participant.

%\subsection{@\codegl}
\revised{\textbf{\textit{@\codegl. }}}As depicted in the left view in \figref{fig:screen_prompt}, a developer can ask Copilot to generate a code snippet for a specific task directly in its chat by prefixing the prompt with \texttt{@\codegl}. The specification of the participant will trigger the \toolName extension. In this example, the developer asks to generate a code snippet to ``extract zip file recursively''.
Once the LLM output is received, \toolName asks the LLM to search for relevant links related to the previously generated code and will process their contents to come up with a set of links that are likely the origin of the generated code. Depending on the similarity thresholds configured by the developer, the set of initial retrieved links is filtered out from noisy data, \eg links related to the coding tasks but without any reference to the automatically generated code snippets. Furthermore, for each link likely related to the generated code, \toolName will identify the license associated with the code snippet to let the developer be aware of how the code could be reused and/or redistributed. During this activity, there is no need for further interactions with the user while the process runs in the background.

The right view in \figref{fig:screen_prompt} shows the visualization of the links likely being the origin for the generated code from \toolName via an HTML panel embedded in the IDE. Specifically, for each link surviving the filter, the panel reports the URL and the License (if any). Since there might be errors in the identification of the license, \toolName only provides a suggestion for developers, \ie \textit{Possible license found:``$<<$LICENSE$>>$''}. 

\revised{\textbf{\textit{@\links. }}}\toolName can also be used when the automatically generated code is created directly within the IDE, \eg using Copilot's code completion. Note that this operational mode can be used for any code snippet in the codebase.  In this case, the developer manually selects a code snippet in the VS Code editor and activates the dedicated command, \textbf{Send Selected Code to \toolName}, through the extension interface.
%, as shown in \figref{fig:screen_menuVSC}.

Once activated, the extension invokes the \texttt{@\links} participant, which automatically constructs a request (described in Section~\ref{sec:architecture}) to the LLM to retrieve links to potential sources that contain code similar to the selected code. %o this end, the internally generated prompt includes an explicit instruction to perform a targeted web search.
The model's response contains a list of links pointing to potentially relevant web pages. 
From this point onward, the pipeline proceeds as described in the previous operational mode. Also in this case, the likely origins of the code are shown to the developer through an embedded HTML view within the IDE. 

%The final results are presented to the user through an embedded HTML view within the IDE.

\section{Preliminary Evaluation}
\label{sec:eval}
\begin{comment}
\begin{table}[t]
    \caption{Precision of \toolName varying the thresholds}
    \vspace{-2mm}
    \label{tab:precision}
    \centering
    \begin{tabular}{lccccc}
    \hline \hline
    \multicolumn{6}{c}{\textsc{\textbf{CodeSearchNet}}} \\    \hline
    & \textbf{0.5} & \textbf{0.6} & \textbf{0.7} & \textbf{0.8} & \textbf{0.9} \\
    \hline
    \textbf{Cloning Ratio} & 0.88 (1) & 0.88 (1) & 0.94 (1) & 1.00 (1) & 1.00 (1) \\
    \textbf{Textual Similarity} & 0.58 (2) & 0.59 (2) & 0.70 (2) & 0.81 (1) & 0.96 (1) \\
    \hline
    \hline
     \multicolumn{6}{c}{\textsc{\textbf{CoderEval}}} \\ \hline 
    & \textbf{0.5} & \textbf{0.6} & \textbf{0.7} & \textbf{0.8} & \textbf{0.9} \\
    \hline
    \textbf{Cloning Ratio} & 0.90 (1) & 0.94 (1) & 0.94 (1) & 1.00 (1) & 1.00 (0)\\
    \textbf{Textual Similarity} & 0.58 (2) & 0.63 (2) & 0.67 (1) & 0.86 (1) & 0.92 (0) \\
    \hline \hline
    \end{tabular}
    \vspace{-4mm}
\end{table}
\end{comment}

\begin{table}[t]
    \caption{\toolName precision for varying thresholds (in parentheses, the average number of returned links per query)}
    \vspace{-2mm}
    \label{tab:precision}
    \centering
    \begin{tabular}{lccccc}
    \hline \hline
    \multicolumn{5}{c}{\textsc{\textbf{CodeSearchNet}}} \\    \hline
    & \textbf{0.5} & \textbf{0.6} & \textbf{0.7} & \textbf{0.8} \\
    \hline
    \textbf{Cloning Ratio} & 0.88 (1) & 0.88 (1) & 0.94 (1) & 1.00 (1) \\
    \textbf{Textual Similarity} & 0.58 (2) & 0.59 (2) & 0.70 (2) & 0.81 (1) \\
    \hline
    \hline
     \multicolumn{5}{c}{\textsc{\textbf{CoderEval}}} \\ \hline 
    & \textbf{0.5} & \textbf{0.6} & \textbf{0.7} & \textbf{0.8} \\
    \hline
    \textbf{Cloning Ratio} & 0.90 (1) & 0.94 (1) & 0.94 (1) & 1.00 (1) \\
    \textbf{Textual Similarity} & 0.58 (2) & 0.63 (2) & 0.67 (1) & 0.86 (1) \\
    \hline \hline
    \end{tabular}
    \vspace{-4mm}
\end{table}

We have performed a preliminary evaluation of \toolName. Specifically, we evaluated the \toolName's precision in a scenario where, given a code snippet, we asked for the retrieval of relevant links. \revised{Note that we have not evaluated the recall since the complete set of true source links (ground truth) for automatically generated code is typically unknown, mostly because there are no existing benchmarks, \ie labeled dataset, of generated code snippets and their exact source links to compare against.}

For the evaluation, we started with 101 coding tasks (46 Java and 55 Python) in our previous work~\cite{msr2025}, retrieved from the \textsc{CodeSearchNet} dataset~\cite{husain2020codesearchnetchallengeevaluatingstate},  for which at least one relevant link was associated with the automatically generated code. After that, we extended the evaluation, randomly selecting 100 coding tasks (50 for each programming language) from \textsc{CoderEval}~\cite{DBLP:conf/icse/YuSRZZMLLWX24}, a benchmark curated from real-world projects mostly used in the literature to evaluate the effectiveness of LLMs for code generation~\cite{hou2023large}.  

For each task, we first asked Copilot to generate the code. After that, we used the automatically generated code to ask the Copilot chat to retrieve links related to it. To assess the tool's precision, two authors
independently assessed each link in the sample and discussed disagreements. %We computed the Cohen’s k~\cite{cohen1960coefficient} inter-rater agreement, which resulted to be XX (XX).

Since \toolName provides only links that are greater than the user-determined thresholds for the maximum cloning ratio and textual similarity, 
\tabref{tab:precision} reports the precision of \toolName in the two datasets, varying the thresholds for the two metrics in the range [0.5-0.8], together with the average number of links shown to the users in parentheses.  
Although we have no objective means to assess \toolName's recall, such numbers are consistent when considering the tasks from the \textsc{CoderEval} benchmark, where the tool provided 263 links, with only 14 (related to 11 different code snippets) being considered as ``highly relevant'' by the manual annotators.  
For the \textsc{CodeSearchNet} tasks, the tool provided 278 links, of which only 29, belonging to 26 snippets, have been considered ``highly relevant'' to the automatically generated code. 
Looking at the precision of \toolName shown in \tabref{tab:precision}, it is possible to state that, in general, the tool has promising performance, \ie the precision increases when the thresholds increase, while the total number of visualized links decreases. 

Finally, manual annotators also evaluated the tool's ability to identify the appropriate license when reusing a code snippet sourced from a web link. %For this task, the Cohen’s k~\cite{cohen1960coefficient} inter-rater agreement was XX (XX). 
Among the 278 links from the \textsc{CodeSearchNet} tasks, \toolName retrieved a license for 109 links, of which only six were wrongly assigned. For the \textsc{CoderEval} benchmark, instead, the manual annotators deemed 120 of the 131 retrieved licenses as correct. Incorrect detection is often due to the retrieval of a license contained in a web page that, while suitable for the web content, is not the actual source code license. Furthermore, the relatively low percentage of links to which it is possible to associate a license ($\simeq 44\%$) is because very often the retrieved links point to blogs/tutorials from which it is not easy to automatically determine the license. In such cases, it will be the responsibility of the developers to investigate the license and, if any, determine the redistribution criteria.

\section{Conclusions and Future Work}
\label{sec:conc}
This paper describes \toolName, a GitHub Copilot extension that integrates with the VS Code editor. The tool can be used by developers during tasks where they seek assistance from LLMs to generate a code snippet and want to obtain information about the ``trustworthiness'' of the automatically generated code, as well as license information, helping them to assess how the generated code can be reused/redistributed within their code base. 

A preliminary evaluation of \toolName highlights its promising performance. Indeed, even if the number of coding tasks from which the tool can come up with at least one possible link having a code snippet highly similar to the generated one is low, the filtering mechanism implemented with several similarity metrics helps filter out noisy data. 
\revised{Regarding license identification, the identified licenses are generally accurate, although, for several links---e.g., those pointing to some web pages---no clear license could be inferred.}
%For what concerns the license identification, instead, while there are many links for which we are not able to determine the right license, when identified, it is almost correct. 

Future work aims at (i) improving the code search capability of \toolName by combining multiple heuristics, (ii) identifying whether the code on the web was, itself, AI-generated, and (iii) performing an extensive empirical evaluation. 

\balance

\section*{Acknowledgments}
\label{sec:ack}
\revised{This project has been financially supported by the European Union
NEXTGenerationEU project and by the Italian Ministry of University and Research (MUR), a Research Project of Significant National Interest (PRIN) 2022 PNRR, project n. D53D23017310001
entitled ‘Mining Software Repositories for enhanced Software Bills
of Materials (MSR4SBOM)’. Daniele Bifolco is partially funded by
the PNRR DM 118/2023 Italian Grant for Ph.D. scholarships.}

\bibliographystyle{IEEEtranS}
%\balance
\bibliography{bib.bib}

\end{document}